\documentclass[preprint,prl,epsf,aps,showpacs,nofootinbib]{revtex4}
\usepackage{bm}
\usepackage{epsfig}
\input epsf

\newcommand{\nix}[1]{}

\begin{document}

\title{Spin relaxation times of 2D holes from spin sensitive bleaching of inter-subband absorption}

\author{Petra~Schneider$^1$, J.~Kainz$^1$, S.D.~Ganichev$^{1,2}$, V.V.~Bel'kov$^2$,
S.N.~Danilov$^1$, M.M.~Glazov$^2$, L.E.~Golub$^2$, U.~R\"{o}ssler$^1$,
W.~Wegscheider$^1$, D.~Weiss$^1$, D.~Schuh$^3$, and W.~Prettl$^1$}
\address{$^1$~ Fakult\"{a}t
Physik, University of
Regensburg, 93040 Regensburg, Germany\\
$^2$~A.F.~Ioffe Physico-Technical Institute,
194021 St.~Petersburg, Russia\\
$^3$~Walter Schottky Institute, TU Munich, 85748 Garching, Germany}

\date{\today}

\begin{abstract}
We present spin relaxation times of 2D holes obtained by means of spin sensitive bleaching of the absorption of infrared radiation in $p$-type GaAs/AlGaAs quantum wells (QWs). 
It is shown that
the saturation of inter-subband absorption of circularly polarized radiation
is mainly controlled by the spin relaxation time of the holes. The saturation behavior has been determined for different
QW widths and in a wide temperature range with the result that the saturation intensity substantially
decreases with narrowing of the QWs.
Spin relaxation times are derived from the measured saturation intensities by making use of calculated (linear) absorption coefficients
for direct inter-subband transitions. It is shown that spin relaxation is due to the D'yakonov-Perel' mechanism governed by hole-hole scattering. The problem of selection rules is addressed.

\end{abstract}


\maketitle \draft

\newpage

\section{I. Introduction}

The spin degree of freedom of charge carriers in semiconductors,
being of fundamental interest as a dynamic variable, has recently attracted
much attention  in view of its possible role in active
spintronic devices \cite{awschalom:2002}. It is intimately related to the polarization degree-of-freedom of electromagnetic waves via
the selection rules which have been used for optical spin orientation
\cite{pikus:1984}. The spin relaxation times of electrons and holes
in semiconductor quantum well structures have been measured for the first
time in time-resolved photoluminescence experiments
\cite{damen:1991,kohl:1991,tackeuchi:1996,terauchi:1999}. In these
investigations with optical excitations across the band gap, electron-hole pairs
are created and the measured spin relaxation times reflect the particular
situation of a bipolar spin orientation with relaxation processes, in
which the electron-hole exchange process can play the dominant role \cite{bir:1976}. This situation is not the one to be expected
in prospective spintronic devices \cite{datta:1990} which are likely
to operate with one kind of carriers only, spin polarized electrons or
holes, injected into the semiconductor via ferromagnetic contacts. For
this situation the monopolar spin relaxation is the decisive dynamical
quantity, whose dependence on device parameters needs to be
investigated.

In spite of recent progress, the injection of spin polarized
carriers through heterocontacts remains a challenge and does not
allow to measure spin relaxation times yet \cite{ohno:1999,schmidt:2000}. Therefore,  monopolar optical
spin orientation combined with the photogalvanic effects (PGE), which has been
demonstrated for $n$- and $p$-doped quantum well structures of different
material compositions \cite{APL00,PRL01}, is the method of
choice to investigate the spin dynamics of electrons or holes 
avoiding the problems connected with electrical spin injection. It has been
demonstrated \cite{PRL02}, that the linear and the circular PGE show a
distinct saturation behavior with increasing intensity of the
exciting light which carries information about the spin
relaxation time. The analysis of these data requires the knowledge of the
linear absorption coefficient for inter-subband transitions, which is
difficult to measure and is hence provided by realistic calculations within the
self-consistent multiband envelope function approximation
\cite{winkler:1993}. 

We present here a detailed investigation of spin
relaxation in rectangular $p$-type (113)-grown GaAs/AlGaAs quantum wells of
different widths $L_W$ and in a wide temperature range. This first comprehensive
experimental study of monopolar spin relaxation in dependence on these two
relevant system parameters, width and temperature,  is accompanied by a theoretical analysis
that relates the measured spin relaxation times to  the D'yakonov-Perel'
mechanism.

The paper is organized as follows. First, we will present our samples and experimental technique and the results of the measurements. Following that, we outline the calculation of the absorption coefficient and by making use of this calculation derive the spin relaxation times. This is followed by a discussion of the dominant spin relaxation mechanism and the topic of selection rules.\\

\section{II. Experiment}

The experiments have been carried out on $p$-type (113) MBE-grown GaAs/AlGaAs QWs  with widths $L_W$ of 7, 10 and 15~nm. In order to improve sensitivity, multiple structures of 20 QWs were investigated.   
Samples with free carrier sheet densities $p_s$ of about $2\cdot 10^{11}$ cm$^{-2}$ and a high mobility $\mu$ of around
$5\cdot10^5\,\mathrm{cm}^2/(\mathrm{Vs})$ (at 4.2 K) were studied in the range from liquid helium temperature up to 140 K. At the samples a pair of Ohmic contacts is centered on opposite sample edges along the direction $x\,\,||\,\,[1\bar{1}0]$.
 As source of radiation a high power pulsed far-infrared (FIR) molecular laser,
optically pumped by a TEA-CO$_2$ laser, has been used delivering 100~ns pulses with intensities up to 1~MW/cm$^2$ in the
wavelength range between 76~$\mu$m and 148~$\mu$m providing direct inter-subband transitions from the lowest heavy hole $hh1$ to the light hole $lh1$ subband. The radiation of the FIR-laser is linearly polarized and a $\lambda$/4 plate was used to generate circularly polarized radiation with a polarization degree $P_{\rm{circ}} = \pm 1$ for right- and left-handed circularly polarized light.\\

The absorption of terahertz radiation
by free carriers in
 QWs is weak due to their small thickness and  difficult to measure in   transmission experiments. This is even worse
in the case of bleaching at high power levels.
Therefore, the  nonlinear behavior of the absorption
has been investigated employing
the recently observed circular (CPGE) and linear (LPGE) photogalvanic
effects \cite{PRL01,APL00}. Both,
CPGE and LPGE
yield an easily
measurable electrical current
in $x$-direction. According to Ivchenko and Pikus \cite{book}
the nonlinear absorption coefficient is proportional to
the photogalvanic current $j_x$
normalized by the
radiation intensity $I$. Thus, by choosing the degree of
polarization, we
obtain a photoresponse corresponding to the absorption coefficient of circularly
or linearly polarized radiation.\\

\begin{figure}[t]
\centerline{\epsfxsize 100mm \epsfbox{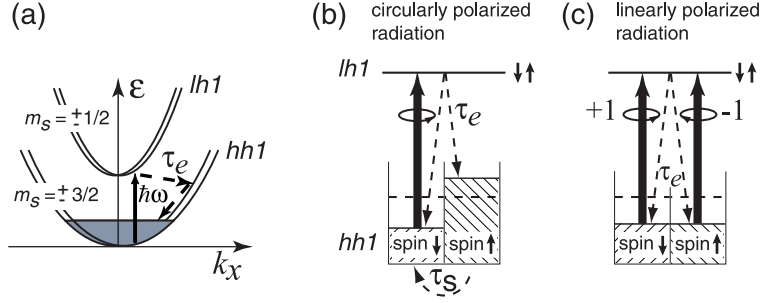}}
\vspace*{-0.3cm}
\caption{Microscopic picture of spin sensitive bleaching: (a)
direct $hh${\it1}-$lh${\it 1} optical transitions, (b) and (c)
process of bleaching for two polarizations. Dashed arrows indicate energy
($\tau_e$) and spin ($\tau_s$) relaxation.}
\label{fig1}
\end{figure}

\begin{figure}[h]
\centerline{\epsfxsize 100mm \epsfbox{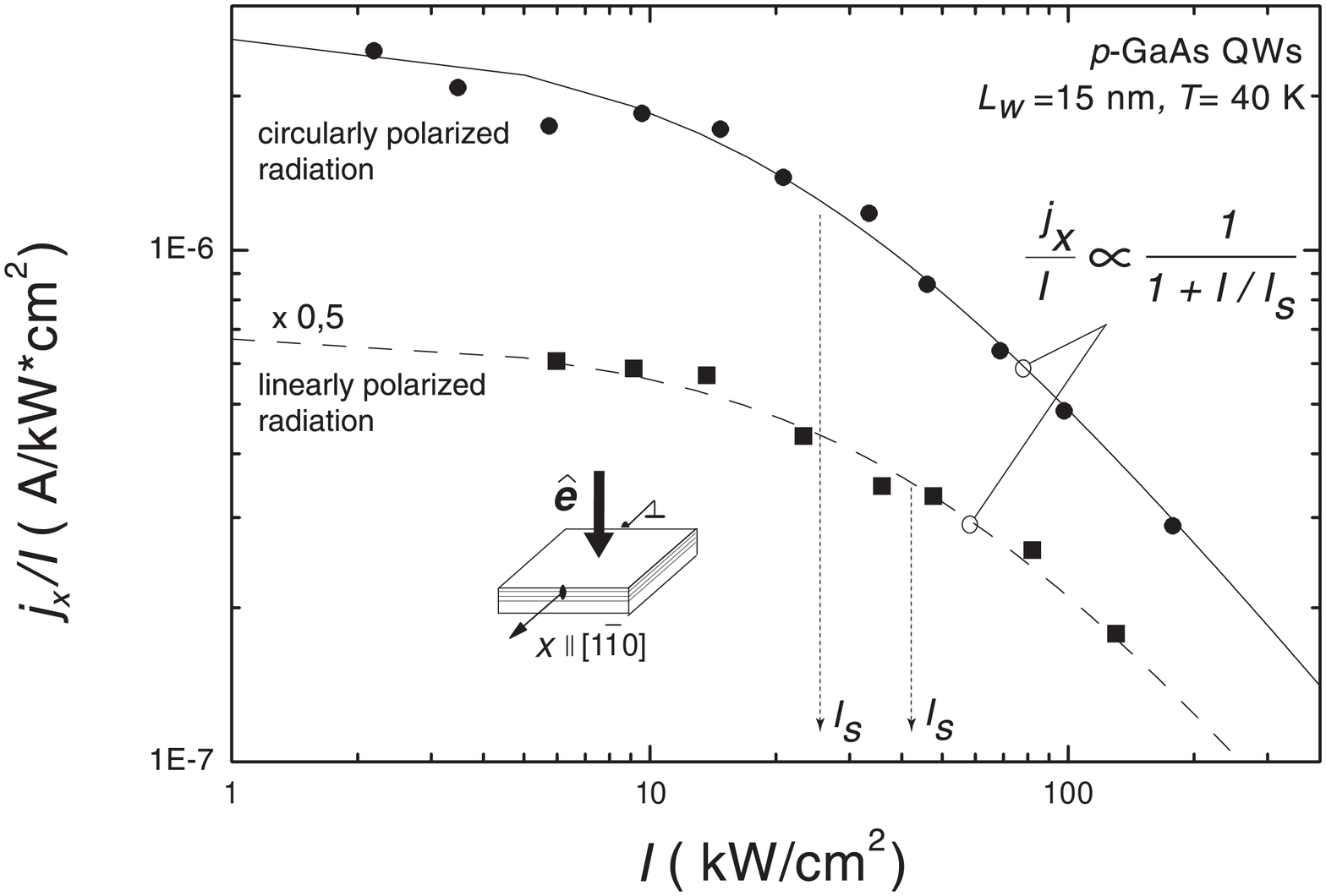}}
\vspace*{-0.2cm}
\caption{CPGE and LPGE  currents $j_x$   normalized by the intensity as a function of the intensity
for  circularly and linearly polarized radiation of $\lambda=148$~$\mu$m and at T = 40 K.
}
\label{fig2}
\end{figure}

The investigated intensity dependence of the absorption coefficient $\alpha\propto j_x/I$ shows saturation with higher intensities for all samples used in our experiments. It is observed that saturation takes place for excitation with circularly polarized radiation  at a lower level of intensity compared to the excitation with linearly polarized radiation. The basic physics of this spin sensitive bleaching of
absorption can be understood by looking at Fig.\,\ref{fig1}. Illuminating a $p$-type sample
with FIR radiation of  an appropriate wavelength 
results in direct transitions between the
heavy-hole $hh${\it 1}  and the light-hole  $lh${\it 1} subbands. This
process depopulates and populates selectively
spin states in
$hh${\it 1} and $lh${\it 1} subbands. The absorption is
proportional to the difference of populations of the initial and
final states. At high intensities the absorption decreases since
the photoexcitation rate becomes comparable to the non-radiative
relaxation rate back to the initial state.
For C$_s$-symmetry, relevant for our (113)-grown QWs, the selection rules for the absorption at $\bm{k}$ close to zero are so that only one type of spin is involved in the absorption of circularly
polarized light (a closer look at selection rules will be given at the end of this paper). Thus the absorption bleaching of circularly
polarized radiation is governed by energy relaxation of
photoexcited carriers and  spin relaxation within the initial spin-split
subband~(see Figs.~\ref{fig1}a and~\ref{fig1}b). These processes are
characterized by  energy and  spin relaxation times $\tau_e$ and
$\tau_s$, respectively. We note that during energy relaxation to
the initial state in  $hh{\it 1}$, the holes loose their photoinduced spin
orientation due to rapid relaxation~\cite{Ferreira91p9687}. Thus,
spin orientation occurs in the initial subband $hh{\it 1}$, only. In
contrast to circularly polarized light, absorption of linearly
polarized light is not spin
selective and the saturation is
controlled by the energy relaxation only~(see Fig.~\ref{fig1}c).
For $\tau_s > \tau_e$, bleaching of absorption becomes spin sensitive and
the saturation intensity $I_s$ of circularly polarized radiation drops
below the value of linear polarization as indicated in  Fig.~\ref{fig2} by arrows. The saturation intensity is defined as the intensity at which $j_x/I$ is one half of its unsaturated value at $I\rightarrow 0$.\\

\begin{figure}[t]
\centerline{\epsfxsize 180mm \epsfbox{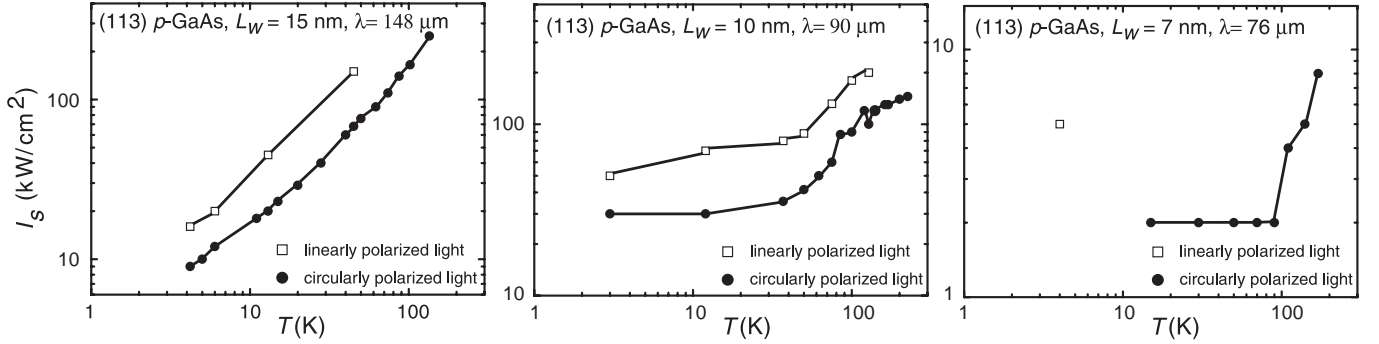}}
\vspace*{-0.2cm}
\caption{Temperature dependence of the saturation intensities for various QW widths for
linearly (open squares) and circularly (full circles) polarized light. The thickness of the QWs decreases from left to right.}
\label{fig3}
\end{figure}

Fig.~\ref{fig3} presents the saturation intensities for different QW widths in the whole investigated temperature range. Note that the saturation intensities $I_s$ for the excitation with circularly
polarized radiation (circles) are generally smaller than for linearly
polarized radiation (squares). A significant reduction
of the saturation intensities with decreasing $L_W$ is observed and indicates longer hole spin relaxation times
for narrower QWs, which has been shown theoretically in \cite{Ferreira91p9687} for the first time. \\

The nonlinear
behavior of the photogalvanic currents has been analyzed in terms of
excitation-relaxation kinetics taking into account both optical
excitation and non-radiative relaxation processes. It was
shown~\cite{PRL02} that the photocurrent $j_{LPGE}$, induced by linearly
polarized radiation, is given by $j_{LPGE}/I \propto (1 +
I/I_{se})^{-1}$, where $I_{se}$ is the saturation intensity
controlled by energy relaxation of the hole gas, while the photocurrent
induced by  circularly polarized radiation
$j_{CPGE} \propto I / \left( 1 + I \left( {I_{se}}^{-1} + {I_{ss}}^{-1}
\right) \right) $
in addition is controlled by spin relaxation with the term $I_{ss}= \hbar\omega p_s /(\alpha_0 L_W
\tau_s)$. Here $\alpha_0$ is the unsaturated absorption coefficient at low
intensities. Thus the spin relaxation time $\tau_s$ is given by

\begin{equation}
\tau_s= \frac{ \hbar \omega p_s}{\alpha_0 L_W I_{ss} }
\label{tau1}
\end{equation}

\section{III. Absorption coefficient}

In order to obtain $\tau_s$ with this formula from the measured saturation intensities $I_{ss}$, the value of $\alpha_0$, not available from experiment, is determined theoretically.
The calculations of the linear absorption coefficient $\alpha_0$ for inter-subband transitions
are based on the self-consistent multiband envelope function approximation
(EFA) \cite{winkler:1993}, that takes into account the crystallographic orientation of
the QW (here the [113] direction) and the doping profile \footnote{ In accordance with the
growth parameters of the samples, we assumed an acceptor concentration of
$1 \cdot 10^{16} \, \mathrm{cm}^{-3}$ in the barriers and a spacer width
of $70 \,$\AA \,\,($45\,$\AA) on the left (right) side of the well. The values
of the band parameters are identical with those given in: L. Wissinger, U.
R\"ossler, B. Jusserand, and D. Richards, Phys. Rev. B {\bfseries 58}, 15375
(1998).}. Calculations are performed
within the Luttinger model of the heavy and light hole states to obtain the hole subband
dispersion $\epsilon_i(\bm{k})$ and eigenstates $|i,\bm{k} \rangle$
of the hole subband $i$ and in-plane wave-vector $\bm{k}$.
For direct (electrical dipole) transitions between subbands $i$ and $j$ the contribution to the
absorption coefficient $\alpha_{i \rightarrow j}(\omega)$ as a function of the
excitation energy $\hbar \omega$ is then given by~\cite{Vorobjev96p981}

\begin{equation}
\hspace*{-0.5cm}
\alpha_{i \rightarrow j}(\omega) = \frac{e^2}{4 \pi \epsilon_0 \omega c n L_W}
 \int \mathrm{d}^2 k \left| \langle j, \bm{k} |
 \bm{e} \cdot \hat{\bm{v}}(\bm{k}) | i, \bm{k} \rangle \right|^2
 \left[ f_j(\bm{k}) - f_i(\bm{k}) \right]
 \frac{
 \mathrm{e}^{-\left(\epsilon_j(\bm{k}) - \epsilon_i(\bm{k}) - \hbar \omega \right)^2
 /\Gamma^2}
 }{\sqrt{\pi} \Gamma}
 \; , \;
\end{equation}

\noindent
where $\bm{e}$ is the light polarization vector, $n$ is the refractive index,
$\epsilon_0$ is the free-space permittivity, $f_i(\bm{k})$ is the
Fermi distribution function in the subband $i$ and $\Gamma$ is a phenomenological parameter
to account for the level broadening due to scattering.
Within EFA, the velocity $\hat{\bm{v}}(\bm{k}) $ is a matrix operator expressed as the gradient
in $\bm{k}$-space of the Luttinger Hamiltonian. Its matrix elements
are calculated from the EFA wave functions.

\begin{figure}[t]
\vspace*{-0.4cm}
\centerline{\epsfxsize 170mm \epsfysize 45mm  \epsfbox{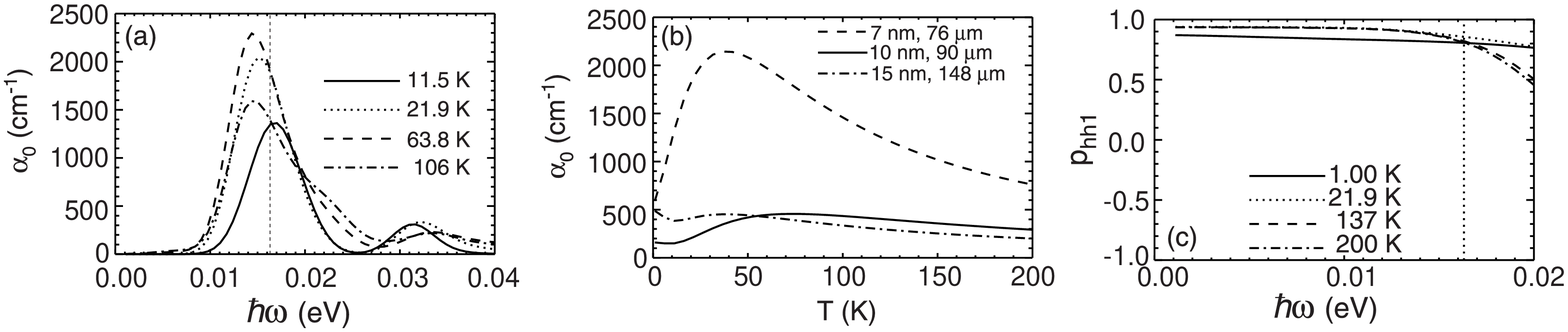}}

\caption{
(a) Calculated absorption coefficient $\alpha_0$ for a QW with $L_W$= 7 nm as a function of photon
energy $\hbar \omega$ for various temperatures $T$ and (b) as a function of {\it T} for various QW widths with $\hbar \omega$ corresponding to the energy of the exciting laser light. (c) Hole spin orientation efficiency $p_{hh1}$ (see chapter {\it selection rules and spin orientation}) as a function of $\hbar \omega$ for different $T$, $L_W$= 7 nm and right handed circular polarization. All calculations were performed for a
carrier density $p_s$ of about $2 \cdot 10^{11}  \, \mathrm{cm}^{-2}$ and a broadening $\Gamma$ = $2.47\, \mathrm{meV}$. 
}
\label{fig_abskoeff}
\vspace*{-0.2cm}
\end{figure}

Following this scheme we calculate the absorption coefficient
$\alpha_0(\omega) = \sum_{ij} \alpha_{i \rightarrow j}(\omega) $.
The absorption spectrum for the system with $L_W=7\,\mathrm{nm}$
is shown in Fig. \ref{fig_abskoeff}a.
At low temperatures two pronounced peaks evolve, which correspond to the transitions
from the lowest (spin split) hole subband to the second and third subband, respectively.
Fig.~\ref{fig_abskoeff}b shows the temperature dependence (due to the Fermi distribution function)
of $\alpha_0$ at the respective excitation
energies for the different samples.
The calculated values of $\alpha_0$ are used to convert the measured saturation intensities $I_{ss}$ according to Eq.~(\ref{tau1}) into spin relaxation times $\tau_s$.\\

\begin{figure}[h]
\centerline{\epsfxsize 90mm \epsfbox{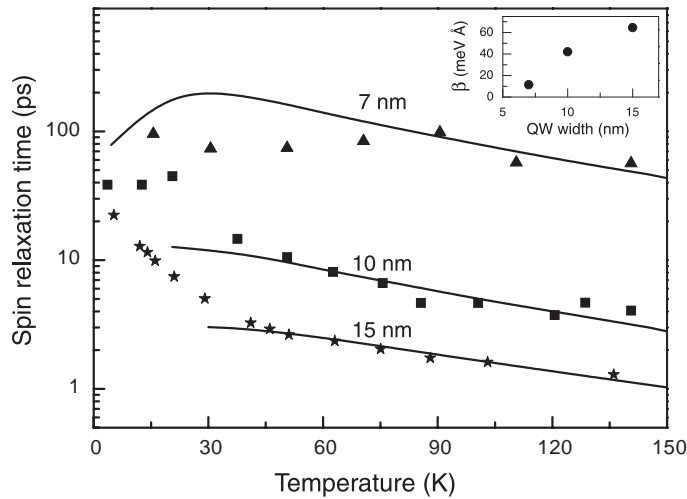}}
\vspace*{-0.2cm}
\caption{Spin relaxation times of holes for three different widths of (113)-grown GaAs/AlGaAs QWs as a function of temperature. The solid lines show a fit according to the D'yakonov-Perel' relaxation mechanism. The inset shows the hole spin-splitting parameter $\beta$ obtained from the fit.
}
\label{fig5}
\end{figure}

The resulting hole spin relaxation times in dependence on the temperature are shown in Fig.~\ref{fig5} for QWs of different widths. Our measurements show longer hole spin relaxation times for narrower QWs. Note the different behavior of the spin relaxation times with the temperature for different QW widths. It is worth mentioning that at high temperatures a doubling of the QW width decreases $\tau_s$ by almost two orders of magnitude. Compared to the values given in \cite{PRL02} (for $L_W$~=~15~nm), where $\alpha_0$ was derived from~\cite{Vorobjev96p981}, we obtain here smaller $\tau_s$
at higher temperatures due to a more realistic theoretical
model for the calculation of $\alpha_0$.\\

\section{IV. Spin relaxation mechanism}

\begin{table}
\caption{Momentum relaxation times $\tau_p$ (determined from mobility) and the ratios $\tau_p / \tau_s$ for different QW widths at 4.2~K.}
\begin{center}
\begin{tabular}{|c|c|c|}
  \hline
  QW width (nm) & $\tau_p$ (ps) & $\tau_p / \tau_s$ \\ \hline
  7 & 9.5 & 0.1 \\ \hline
  10 & 25 & 0.64 \\ \hline
  15 & 38 & 1.73 \\ \hline
  \end{tabular}
  \end{center}
\label{table3}
\end{table}

In order to understand the mechanism governing spin relaxation, we
consider the ratio of momentum $\tau_p$ and spin $\tau_s$
relaxation times at $T=4.2$~K presented in Table~\ref{table3}.
In the $p$-doped QWs, studied here, there are two possible routes to
hole spin rela\-xation: the Elliot-Yafet and the
D'yakonov-Perel' mechanism. In the first case, spin is lost during
scattering. However the ratio $\tau_p / \tau_s$ for holes, where $\tau_p$ is determined from mobility measurements, has a
strong dependence on the QW width ($\sim L_W^6$) for scattering from impurity or
interface micro-roughness. 
Note that for the calculation of the spin relaxation time we do not take into account phonon scattering because most of the experimental data belongs to the range of low temperatures where phonon scattering processes play an unimportant role.
 In
addition, $\tau_p$ is of the same order as $\tau_s$ for the two
wider QWs which contradicts the main idea of the Elliot-Yafet
mechanism. Another possibility is the Elliot-Yafet spin
relaxation controlled by hole-hole collisions, but for this
mechanism asymmetry of the QW heteropotential is
needed~\cite{glazovHoles}.

We conclude that the
Elliot-Yafet mechanism is unimportant in the structures under
study, since the experiment shows a too weak dependence for $\tau_p /
\tau_s$ on the QW width. The above  experimental results suggest much longer spin relaxation times for the given mobilities than expected for the  Elliot-Yafet mechanism. The spin relaxation time at helium temperature
according to Elliot-Yafet mechanism can be estimated as 
$$\tau_s \approx
\tau_p \left( \frac{k_{\rm F} L_W}{\pi} \right)^{-6}$$

\noindent
where $\bm{k}_{\rm F}$ is the Fermi wave-vector.
 This yields
$\tau_s \approx 5\times 10^5$~ps which is three orders of magnitude larger than measured values. Therefore, the main mechanism of hole spin relaxation is the
D'yakonov-Perel' one~\cite{Bastard93p439}: hole spin is lost between the scattering
events. For this mechanism, the spin relaxation rate is given by the expression

\begin{equation}\label{tausDP}
    {1 \over \tau_s} = \left(\beta\over\hbar\right)^2 \: k_{\rm F}^2 \: \tau^*,
\end{equation}

\noindent
where  $\beta$ is the spin-splitting coefficient of the $\bm k$-linear terms in the Hamiltonian yielding 
\[
E_{3/2}(k) - E_{-3/2}(k) = 2 \: \beta \: k.
\]

The time $\tau^*$ is the microscopic scattering time which has
contributions from both momentum scattering and carrier-carrier
collisions~\cite{glazov02}. We have calculated the hole-hole
scattering time governing the D'yakonov-Perel' spin relaxation
mechanism by sol\-ving the quantum kinetic equation for the hole
pseudospin density matrix similar to Ref.~[\onlinecite{glazov03}].
Our calculation shows that the hole-hole scattering time is shorter
than $\tau_p$ at 4.2~K. We believe that in the relevant
temperature range $\tau_p$ does not change significantly.
Therefore, hole-hole scattering controls D'yakonov-Perel' spin
relaxation in the whole temperature range.

Fig.~\ref{fig5} presents spin relaxation times extracted from the 
experiment (points) together with a theoretical fit using Eq.~(\ref{tausDP})
(solid lines) showing  a  good agreement between theory and
experiment. The discrepancy at low lattice temperatures may be
attributed to the fact that the hole gas is not in equilibrium due
to optical pumping. This case requires special theoretical
treatment.

In the inset the hole spin-splitting parameter $\beta$ obtained
from the fit is plotted as a function of the QW width. The
corresponding spin splitting is equal to 0.17, 0.68, and 1.32~meV
for QW widths 7, 10, and 15~nm, respectively. This order of
magnitude agrees with hole spin-splitting obtained from multiband
calculations~\cite{winkler00}. The parameter $\beta$
increases with the QW width. This is a specific feature of 2D hole
systems where spin-splitting is determined by heavy-light hole
mixing which is stronger in wider QWs~\cite{winkler02}.\\

\section{V. Selection rules and spin orientation}

For the definition of $I_{ss}$ we assumed that the spin
selection rules are fully satisfied at the transition energy.
This is the case for optical transitions occurring close to $\bm{k}=0$
in (001)-grown systems~\cite{Jorda90p481}.
However, in (113)-grown systems, heavy-hole and light-hole subbands  are strongly mixed, even at $\bm{k}=0$.
This reduces the strength of the selection rules
and therefore the
efficiency of spin orientation. The mixing can be taken into account
by means of a multiplicative factor in $I_{ss}$, which increases the saturation
intensity at constant spin relaxation time~\cite{PhysicaE01}.

The lowest subband, which for (001)-grown systems is purely heavy hole ($m_s=\pm 3/2$) 
at $\bm{k}=0$, has for growth direction [113] an admixture of
about $10 \%$ 
light hole spinor components ($m_s=\pm 1/2$) \cite{winkler:1996}. 
This admixture is sufficiently small to justify subband labeling according to the dominant spinor component at $\bm{k}=0$.

Strict selection rules for inter-subband transitions between hole subbands only exist for some idealized limits
(e.g. spherical approximation for the Luttinger Hamiltonian or growth directions of high symmetry and $\bm{k}=0$).
However,
assuming a symmetrically doped (113)-grown QW, the lowest $hh$ and $lh$ subband states ($hh{\it 1}$ and $lh {\it 1}$, respectively) 
have even parity at $\bm{k}=0$
and no transition between $hh {\it 1}$ and $lh {\it 1}$ is possible, as the velocity operator projected on the light 
polarization direction $\hat{\bm{v}} \cdot \bm{e}$ 
couples only states of different parity. Therefore a strictly valid selection rule cannot be obtained and a more
quantitative discussion of the relative weight of the possible transitions is necessary.
For $\bm{k}$ small enough to ensure that the admixture of odd parity spinor components is negligible, only contributions 
in $\hat{\bm{v}} \cdot \bm{e}$ linear in $\bm{k}$ are to be considered.

A more detailed analysis gives the following results:
The spin-conserving transitions 
$hh {\it 1} \uparrow \, \rightarrow  \, lh{\it 1} \uparrow$ and $hh {\it}1\downarrow \, \rightarrow \, lh {\it1}\downarrow$ are much
weaker than the corresponding spin-flip transitions
$hh {\it 1} \uparrow \, \rightarrow \, lh {\it 1} \downarrow$ and $hh {\it 1} \downarrow \, \rightarrow \, lh {\it 1} \uparrow$.
Depending on the left/right circular polarization of the exciting light, one of the spin-flip transitions is 
dominant.
To investigate the hole spin orientation, we also performed a numerical calculation 
of $\alpha_{i \rightarrow j}$ for excitation with right-hand circularly polarized light.
We obtained that the transition $hh {\it 1} \downarrow \, \rightarrow \, lh {\it 1} \uparrow$ is far more probable 
than all other transitions.
This is quantitatively described by the heavy hole spin polarization efficiency
\begin{equation}
p_{hh1} = \frac{ \sum_i \alpha_{hh1\downarrow \rightarrow i}- \alpha_{hh1\uparrow \rightarrow i} }{ \sum_i
\alpha_{hh1\downarrow \rightarrow i} + \alpha_{hh1\uparrow \rightarrow i} }
\; ,
\end{equation}
where the summation is performed over all subbands.
If $p_{hh1}$ is $+ 1$ ($-1$) the excitation leaves only heavy holes belonging to the up (down) branch of the dispersion
in the $hh {\it 1}$ subband.
In our case, $p_{hh1} $
is around 80 \% at the laser excitation energy and almost independent of the temperature (Fig.~\ref{fig_abskoeff}c).
Therefore we may neglect effects due to incomplete spin orientation, as assumed in this contribution.\\

In conclusion our experimental results demonstrate a strong dependence of the hole spin relaxation times on the width of the quantum well. With wider QWs, the spin relaxation times become much shorter. At high temperatures, a doubling of the QW width results in a changing of the magnitude in two orders. Theoretical calculations in comparison to quantitative experimental results show that the D'yakonov-Perel' mechanism controlled by hole-hole collisions dominates the spin relaxation process.\\

The authors thank E. L. Ivchenko for helpful discussions and fruitful comments.
Financial support from the DFG, the RFBR, ``Dynasty'' Foundation --- ICFPM and INTAS is gratefully acknowledged.

\end{document}